\documentclass[aip,cha,amsmath,amssymb,reprint]{revtex4-1}

\usepackage{graphicx}
\usepackage{dcolumn}
\usepackage{color}
\usepackage{ulem}

\begin{document}

\title{Citrate synthase proteins in extremophilic organisms -- studies within
a structure-based model}

\author{Bartosz R\'{o}\.{z}ycki}
\email{rozycki@ifpan.edu.pl}
\affiliation{Institute of Physics, Polish Academy of Sciences,\\
Al. Lotnik{\'o}w 32/46, 02-668 Warsaw, Poland}

\author{Marek Cieplak}
\affiliation{Institute of Physics, Polish Academy of Sciences,\\
Al. Lotnik{\'o}w 32/46, 02-668 Warsaw, Poland}

\date{\today}

\begin{abstract}
\noindent 
We study four citrate synthase homodimeric proteins within a structure-based 
coarse-grained model. Two of these proteins come from thermophilic bacteria, 
one from a cryophilic bacterium and one from a mesophilic organism; three
are in the closed and two in the open conformations. Even though the
proteins belong to the same fold, the model distinguishes
the properties of these proteins in a way which is consistent with experiments.
For instance, the thermophilic proteins are more stable 
thermodynamically than their mesophilic and cryophilic homologues, 
which we observe both in the magnitude of thermal fluctuations near 
the native state and in the kinetics of thermal unfolding. 
The level of stability correlates with the average coordination number
for amino acids contacts and with the degree of structural compactness. 
The pattern of positional fluctuations along the sequence in the closed conformation 
is different than in the open conformation, including within the active site. 
The modes of correlated and anticorrelated movements of pairs of amino acids 
forming the active site are very different in the open and closed conformations. 
Taken together, our results show that the precise location of 
amino acid contacts in the native structure appears to be a critical element 
in explaining the similarities and differences in the thermodynamic properties, 
local flexibility and collective motions of the different forms of the enzyme. 
\end{abstract}

\keywords{proteins, native structure, citrate synthase, thermal stability, fluctuations, 
coarse-grained models, structure-based models, molecular dynamics simulations}

\maketitle

\section{Introduction}

The primary characteristics of globular proteins are their
native structures and sequences of amino acids.
An evolutionary divergence from
a common ancestor may lead diverse sequences to fold to 
nearly the same native structure. One particularly interesting
example is citrate synthase (CS), an enzyme that is found in most living  
organisms, from bacteria to man \cite{Labe_book}. It acts as a part of
the Krebs cycle that generates ATP \cite{Krebs}.
The native structure of CS from a bacteria living in hot hydrothermal vents
is superimposable with that of CS from a bacteria living under the 
conditions of extreme cold \cite{Labe_book} 
and yet properties of these proteins, such as the thermodynamic stability, 
are distinct so that the enzymatic action 
-- conversion of acetyl-CoA and oxaloacetate into CoA and citrate \cite{Lehninger} -- 
is executed properly.

In this paper, we investigate to what extent structure-based 
coarse-grained models, or rather their one specific implementation, 
can capture physical differences between proteins which belong to the same fold.
We study four CS proteins and demonstrate that their properties are indeed 
distinct in the model due to the existence of slight differences in 
the native structures. 
In particular, the model thermophilic proteins are found to have 
stronger thermodynamic stability than those which are cryophilic 
(an equivalent term is: psychrophilic). 
However, we also find that the root-mean-square 
single-site fluctuations are alike within the very region of the
active site of the thermophilic and cryophilic forms,
even though they are distinct in other parts of the proteins. 

The structure-based dynamical models follow from Go's idea \cite{Ueda,Go,Go0}
that kinetic and thermodynamic properties of proteins should depend primarily
on the geometry of the native structure
and less so directly on the specificity of the sequence. 
There are many variants of such models (see, for instance, Refs.
\cite{Hoang,Clementi2000,Karanicolas,Levy,models,Noel}).
They are not equivalent, but their predictions are expected to
be most accurate
in a vicinity of the native state -- a situation  encountered, for
instance, during stretching manipulations \cite{JPCM,plos,Proteins2014,Current}. 
It has also been argued
that protein folding processes may be proceeding as in the Go-like
models due to minimization of hindrances to folding
\cite{Wolynes,Onuchic,Alm,Takada,Micheletti,Baker}.
Large-scale conformational changes in proteins can be explored by multi-state
Go-type models \cite{Best2005,Chen2007} and mixed elastic network models \cite{Zhu2009}.
In contrast, standard elastic network models, which are less demanding 
computationally than molecular dynamics simulations, are known to capture 
local fluctuations about the native state \cite{Erman1997}. 
Here, we employ a standard Go-type model to examine thermal stability, 
local flexibility, collective motions and unfolding kinetics of CS enzymes 
functioning in organisms that live in very different environmental conditions. 

The adaptations of life to different environmental conditions can be observed 
at various levels of organization, including the molecular level. 
The composition and structure of proteins from organisms 
living under extreme conditions \cite{Cowan,Feller,Collins} 
are known to correlate with the character of the environment. 
For instance, thermophilic
organisms, i.e. those thriving between 45$^{\circ}$~C and 120$^{\circ}$~C,
tend to contain proteins with smaller cavities, bigger number of
ionic bonds, increased polarity of exposed surfaces, an increased
content of charged residues and tryptophan, and a smaller content of 
phenylalanine, methionine and asparagine compared to proteins
in mesophilic organisms \cite{Argos,Bohm,Szilagyi,Claverie}.
On the other hand, cryophilic organisms, i.e. those which
grow and reproduce between about -20$^{\circ}$~C to +10$^{\circ}$~C,
tend to contain proteins with larger catalytic cavities, reduced
content of proline and arginine (to make the backbone more
flexible), increased content of clusters of glycines,
less hydrophobic cores (to make the protein less compact),
a higher proportion of non-polar groups on the surface, and  an
increased content of negative charges on the surface (to facilitate
interactions with the solvent) \cite{Gerday,Cassidy,Bjelic2008,Maayer}.
Structure-based models cannot explicitly take all of these many detailed
chemical features into account, but these features lead to a
particular structure of the native conformation. Therefore, such
models can tell the thermophilic and cryophilic proteins
apart even if they belong to the same fold. 
We shall inquire here -- to what extent.
These models also identify properties that are similar. 
We illustrate these aspects here by considering CS.

\begin{figure}[ht]
\begin{center}
\scalebox{0.4}{\includegraphics{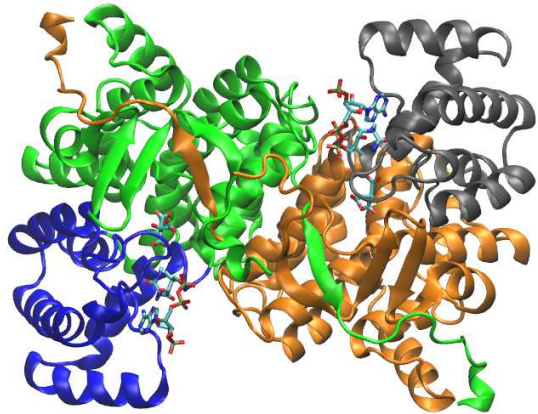}}
\caption{\label{fig:0} Structure of citrate synthase homodimer from 
{\it Pyrococcus furiosus} in the closed conformation (PDB code 1AJ8). 
The reaction products (citrate and CoA) are shown in the stick representation. 
The small and large domain in chain A (first monomer) are shown in blue and green, 
respectively, whereas those in chain B (second monomer) in gray and orange. }
\end{center}
\end{figure}

CS is an $\alpha$-protein homodimer. Fig.~\ref{fig:0} shows
an example of the native conformation -- it is for CS from
{\it Pyrococcus furiosus} with the 
Protein Data Bank (PDB) code 1AJ8 and atomic coordinates
resolved for 741 amino acids \cite{Russell1997}. 
It is formed of two identical subunits, each with its own active site. 
Each of the subunits comprises two domains -- 
a small domain, which comprises five  $\alpha$-helices, and
and a large domain which contains 13 $\alpha$-helices.
In other species, the number of the helices in the large domain
varies between 11 and 15.
The substrate binding site is located in a cleft between the two domains. 

Two kinds  of conformations of CS have been identified: 
one is termed 'open' and the other 'closed'. 
CS is in the 'open' conformation when the active site is not occupied and 
is thus free to bind substrate molecules for the enzyme catalysis. 
When the substrate binds, the small domain undergoes a rotation, 
sealing the substrate binding site and forming the 'closed' conformation. 
A number of crystal structures of CS
from different organisms are available in the PDB. The five  
structures we have considered in our study are summarized in Table~\ref{table:1}. 
It would be ideal to also consider an open form of CS form a cryophilic
organism, but they do not seem to be available in the PDB.

\begin{table}[ht]
\begin{center}
\begin{tabular}{|c|c|c|c|}
\hline
PDB code & confor-  & organism & optimal  \\ 
and & mation & & growth \\
reference & & & temperature \\
\hline
1AJ8 \cite{Russell1997} & closed & {\it Pyrococcus furiosus} & 100$^{\circ}$~C \\
\hline
2CTS \cite{Remington1982} & closed  & {\it Sus scrofa} (pig) & 37$^{\circ}$~C \\
\hline
1A59 \cite{Russell1998} & closed & {\it Antarctic bacterium} & 0$^{\circ}$~C \\
\hline
2IBP \cite{Boutz2007} & open   & {\it Pyrobaculum aerophilum} & 100$^{\circ}$~C \\
\hline
1CTS \cite{Remington1982} & open   & {\it Sus scrofa} (pig) & 37$^{\circ}$~C \\
\hline
\end{tabular}
\end{center}
\caption{\label{table:1} Structures of CS from 
different organisms used in this study. Structures of the closed 
conformation contain the reaction products (CoA and citrate). 
The products are absent in the open conformation structures.}
\end{table}

\section{Methods \label{section:Methods}}

We use a coarse-grained continuum representation for CS configurations 
in which only the positions of C$_{\alpha}$ atoms are retained in 
the molecular dynamics simulations. However, the contact map, i.e.
a list of non-bonded interactions is determined based on the all-atom coordinates.
The configurational energy of the model reads \cite{Hoang,Clementi2000,Hoang2003}
\begin{equation}
\begin{split}
E (X, X_0) & = \sum_{\rm bonds} K_r \left( r - r_{0} \right)^2
+ \sum_{\rm angles} K_{\theta} \left( \theta - \theta_{0} \right)^2 \\
& + \sum_{\rm dihedral} \left[ 
K_{\phi}^{(1)} \left[ 1 - \cos \left( \phi - \phi_{0} \right) \right] \right. \\
& \left. + K_{\phi}^{(3)} \left[ 1 - \cos \left( 3 \phi - 3 \phi_{0} \right) \right] \right] \\
& + \sum_{ {\rm NC} \; (|i-j|>3)} \epsilon \left[ 
5 \left( \frac{r_{0ij}}{r_{ij}} \right)^{12} -
6 \left( \frac{r_{0ij}}{r_{ij}} \right)^{10} \right] \\
& + \sum_{{\rm NNC} \; (|i-j|>3)} \epsilon \left( \frac{\rho_{0}}{r_{ij}} \right)^{12}
\label{eq:Hamiltonian}
\end{split}
\end{equation}
Here, $r$ and $r_0$ denote the distances between two subsequent residues 
at configurations $X$ and $X_0$, respectively, where 
the reference configuration $X_0$ corresponds to the native state. 
The summation over bonds includes peptide and disulfide  bonds
(the latter are present in 2IBP, one in each monomer).
Analogously, $\phi$ and $\phi_0$ represent the bond angles formed by 
three subsequent residues along the amino acid chain at 
configurations $X$ and $X_0$, respectively. 
Next, $\theta$ and $\theta_0$ are the dihedral angles formed by 
four sequential residues at configurations $X$ and $X_0$.  The
coupling parameters in the first three terms in Eq.~(\ref{eq:Hamiltonian}) 
are taken to be as in Ref.~\cite{Clementi2000}, i.e., 
$K_{r} = 100 \; \epsilon /${\AA}$^2$, $K_{\theta} = 20 \; \epsilon$,
$K_{\phi}^{(1)} = \epsilon$, and $K_{\phi}^{(3)} = \epsilon / 2$. 

The fourth term in Eq.~(\ref{eq:Hamiltonian}) describes the native interactions. 
Here, $r_{ij}$ and $r_{0ij}$ denote the distances at configurations $X$ and $X_0$, 
respectively, between residues $i$ and $j$ forming the native contacts. 
The native contacts are identified as in Ref.~\cite{JPCM,plos,Hoang2003}, 
i.e., by using an overlap criterion \cite{Tsai} applied to the coordinates of 
all heavy atoms in the native structure. 
The heavy atoms are assigned enlarged van der Waals spheres and
if there is a pair of atoms for which one finds spheres that
overlap in the native state then the corresponding pair of
amino acids is considered as forming a native contact.
Physically, these contacts are due to hydrogen bonds and ionic bridges.
Only contacts with $|i-j|>3$ are included 
in the contact map.  
The last term in Eq.~(\ref{eq:Hamiltonian}) describes repulsion between non-native 
contacts, with $\rho_0 = 4$~{\AA}. 

The model described by Eq.~(\ref{eq:Hamiltonian}) is different from the
one used in our recent studies of protein stretching \cite{JPCM,plos,Proteins2014}: 
the contact potential is of the 10-12 type, and not 6-12, 
and the local backbone stiffness is given by the usual 
bond-angle and dihedral-angle potentials instead of 
a term involving the local chirality (which effectively 
includes the dihedral term but not the bond angle terms \cite{models}).
The 6-12 model yields amino-acid positional fluctuations 
substantially larger than those found in experiments through 
the temperature factors, and coming close to the experimental data 
requires rescaling as in Ref.~\cite{Prot_MC_2009}. 
Nevertheless, we expect that the calibration of $\epsilon$ 
is of the same order, i.e. $\epsilon \approx 110$ pN\;{\AA}. Thus
the room temperature should be in the vicinity of 0.35 $\epsilon /k_B$,
where $k_B$ denotes the Boltzmann constant.

In our coarse-grained simulations, the citrate and CoA molecules 
(present in the closed structures of CS) 
are represented by one and four beads, respectively. 
The contacts between CS and citrate, 
and between CS and CoA, are found using the same all-atom overlap criterion 
as the native contacts between the amino acids within CS. 
To avoid dissociation of the citrate and CoA molecules from CS during simulations, 
these contacts are replaced by harmonic bonds analogous to the C$_{\alpha}$-C$_{\alpha}$ 
pseudobonds with the spring constant $K_{r} = 100 \; \epsilon /${\AA}$^2$. 

To study thermodynamic properties of the CS dimers, 
we performed overdamped Langevin dynamics simulations 
using in-house software \cite{Hoang2003,Proteins2014}. 
The simulations were carried out at 31 different temperatures $T$ 
distributed uniformly in the interval from 
$0.1 \, \epsilon / k_B$ to $0.7 \, \epsilon / k_B$. 
Each simulation was $10^5 \, \tau$ long and it was preceded
by a $10^4 \, \tau$ equilibration run. The unit of time, $\tau$, is of order 1~ns.
At these temperatures and time scales the dimers never dissociate or unfold. 
We observe only small and moderate deviations from the native state
as measured by RMSD, see Figs.~\ref{fig:1}b and \ref{fig:2}b in the next section. 

In the course of the simulations, we monitor the number of contacts 
\begin{equation}
m (t) = \sum_{i=1}^{M} \theta \left( d_{ij} - r_{ij} (t) \right)
\end{equation}
where $M$ is the total number of contacts in the native state, 
$r_{ij} (t)$ is the distance between residues $i$ and $j$ 
at simulation time $t$, $d_{ij} = 1.2 \, r_{0ij}$ 
is a cutoff distance, and $\theta$ is the the Heaviside function: 
$\theta (x) = 1$ if $x \ge 0$ and $\theta (x) = 0$ if $x<0$. 
To assess thermal stability of the enzyme, we compute 
the probability of finding the enzyme in the native state
\begin{equation}
P_0 = \langle \delta_{m,M} \rangle
\label{P0:eq}
\end{equation}
as a function of temperature $T$, where the brackets denote 
time average after equilibration, and $\delta_{m,M}$ is the Kronecker delta: 
$\delta_{m,M}=1$ if $m=M$ and $\delta_{m,M}=0$ otherwise.
Notice that the definition of $P_0$ involves counting conformations  
in which all native contacts are present. 
The native state probability as given by Eq.~(\ref{P0:eq}) 
is thus different from the average fraction of contacts
\begin{equation}
Q = \langle \frac{m}{M} \rangle
\label{Q:eq}
\end{equation}
We define temperature of thermodynamical stability, 
$T_f$, as one at which $P_0 = \frac{1}{2}$. 
This temperature is different and substantially lower than 
the temperature at which $Q = \frac{1}{2}$. 
Our definition of $T_f$ yields values of $T$ that are in a vicinity 
of temperatures at which smaller proteins fold optimality \cite{Hoang2003}. 

To describe how far a particular configuration has departed from the 
native state, we compute the root mean square deviation
\begin{equation}
{\rm RMSD} (t) = \left[ \frac{1}{N} \sum_{i=1}^{N} \left( \vec{r_{i}} (t) 
- \vec{r_{i}}^{\rm NAT} \right)^2 \right]
\label{eq:RMSD}
\end{equation}
where $\vec{r_{i}}^{\rm NAT}$ denote the positions of C$_{\alpha}$ atoms 
in the native state and $\vec{r_{i}}$ are positions of the C$_{\alpha}$ atoms 
at time $t$ after superimposing on the native structure. 
We use the Kabsch algorithm \cite{Kabsch} to superimpose the instantaneous
structures on the native structure. 
After equilibration, RMSD fluctuates around its average value, 
$\langle {\rm RMSD} \rangle$, which is a function of temperate $T$. 

To quantify the local flexibility of the enzyme, 
we compute the root mean square fluctuation (RMSF) of each residue 
\begin{equation}
\delta_i = \left[ \langle \vec{r_{i}}^{2} \rangle - \langle \vec{r_{i}} \rangle^2 \right]^{1/2}
\label{eq:rmsf}
\end{equation}
Here $\vec{r_{i}}$ is the position of the $i$-th C$_{\alpha}$ atom, 
and the angle brackets denote the average after superimposing on the native structure. 
The RMSFs are directly related to the so called temperature factors \cite{Erman1997} 
\begin{equation}
\beta_i = \frac{8 \pi^2}{3} \delta_{i}^{2}
\label{eq:beta_i}
\end{equation}
which can be determined in biomolecular crystallography experiments
and are listed in PDB files.

The correlation in the motions of residues $i$ and $j$ are given by
\begin{equation}
C_{ij} = \frac{ \langle \delta \vec{r_i} \cdot \delta \vec{r_j} \rangle }
{ \langle \delta \vec{r_i}^2 \rangle^{1/2} \, \langle \delta \vec{r_j}^2 \rangle^{1/2} }
\label{eq:Cij}
\end{equation}
where $\delta \vec{r_i} = \vec{r_i} - \langle \vec{r_i} \rangle$ 
is the displacement from the average position computed 
after superimposing on the native structure. 
Significant positive correlations mean that the residues tend to move together, 
possibly as part of a larger structural motif. Negative
correlations imply that the two residues tend to move in opposite directions. 

To study thermal unfolding and dissociation of the CS dimers, 
we performed overdamped Langevin dynamics simulations 
at elevated temperatures $T$ between 
$1.1 \, \epsilon / k_B$ and $1.7 \, \epsilon / k_B$. 
At each of temperatures we run $200$ trajectories 
to study kinetics of the unfolding and dissociation processes. 

\section{Results}

\subsection{Thermodynamic properties of CS dimers \label{sec:3.1}}

\begin{figure}[ht]
\begin{center}
\scalebox{0.68}{\includegraphics{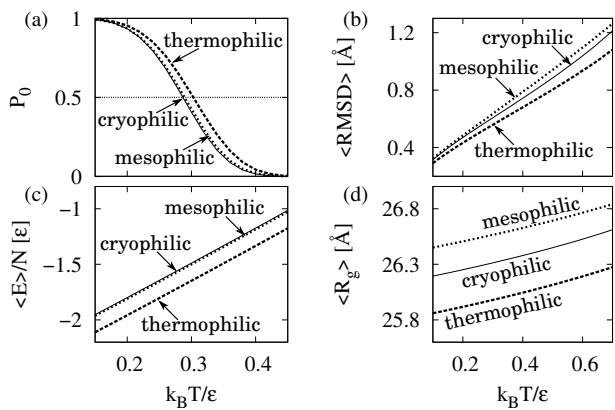}}
\caption{\label{fig:1} 
Thermodynamic properties of citrate synthase from 
thermophilic (1AJ8, dashed lines), mesophilic (2CTS, dotted lines) and 
cryophilic (1A59, thin solid lines) organisms in the closed conformation. 
(a) The probability of finding the enzyme in the native state, $P_0$, 
(b) the average root mean square deviation (RMSD) from the native state, 
(c) the average internal energy per residue, and 
(d) the average radius of gyration, $\langle R_g \rangle$, as functions of temperature $T$. 
}
\end{center}
\end{figure}

We first compare some thermodynamic properties of the CS dimers from 
thermophilic and cryophilic organisms in the closed conformation 
(PDB codes 1AJ8 and 1A59, respectively).
Fig.~\ref{fig:1}a shows that, at any a given $T$, the probability 
$P_0$ of finding the enzyme in the native state is larger for 
the thermophilic CS than for the cryophilic one.  In particular, 
$T_f$ is $0.303 \, \epsilon/k_B$ and $0.287 \, \epsilon/k_B$ 
for the thermophilic and cryophilic CS dimers, respectively. 
With our calibration of $\epsilon$, the difference is of order
14$^{\circ}$~C.

Fig.~\ref{fig:1}b shows that, at any given $T$, $\langle$RMSD$\rangle$ 
is larger for the cryophilic CS than for the thermophilic one. 
Thus the cryophilic CS undergoes larger thermal fluctuations than
the thermophilic CS at the same $T$. 
The internal energy per residue,
shown in Fig~\ref{fig:1}c, is lower by about $0.2 \, \epsilon$ 
for the thermophilic CS compared to the cryophilic CS, 
independent of the value of $T$.
This result is consistent with the data in Fig.\ref{fig:1}a 
showing that the thermophilic CS exhibits better thermal 
stability than the thermophilic CS. 

Interestingly, the thermophilic CS appears slightly more compact 
than the cryophilic CS, as evidenced by the time-averaged radius of
gyration, $\langle R_g \rangle$, plotted against $T$, see Fig.~\ref{fig:1}d 
(note that these two dimers comprise comparable numbers of residues, 
740 and 754, respectively). The difference in $\langle R_g \rangle$ 
is about 0.4~{\AA} or about 1.5\% of the native $R_g$. 
We have checked that, at a given temperature $T$ in the range between 
$0.1 \; \epsilon / k_B$ and $0.7 \; \epsilon / k_B$, 
the standard deviation of the instantaneous $R_g$ values is 
smaller than 0.1~{\AA}, both for the cryophilic and thermophilic forms of CS. 
Thus the observed difference in the values of $\langle R_g \rangle$ 
is small but statistically relevant. 

We next discuss the thermodynamic properties of 
the mesophilic CS in the closed conformation (with the PDB code 2CTS). 
We find that, at any given temperature, $P_0$ 
for the mesophilic CS is smaller than for the thermophilic one 
and only slightly bigger than for the cryophilic CS, see Fig.~\ref{fig:1}a. 
Its stability temperature is $T_f=0.289 \; \epsilon / k_B$. 
The stability difference, as measured by alterations in $T_f$, 
between the thermophilic and mesophilic CS dimers 
in the closed conformation is of order of 12$^{\circ}$~C. 

Interestingly, out of the three CS dimers in 
the closed conformation, the mesophilic one shows the fastest increase 
in $\langle$RMSD$\rangle$ with $T$, see Fig.~\ref{fig:1}b. 
At any given $T$, the internal energy per residue for 
the mesophilic and cryophilic CS dimers in the closed conformation 
are almost the same, see Fig.~\ref{fig:1}c. 
This observation is consistent with the results shown in Fig.\ref{fig:1}a, 
namely, that these two forms are characterized by very similar $P_0 (T)$ curves. 
The mesophilic CS dimer comprises 874 amino acid residues whereas its 
cryophilic and thermophilic homologues -- only 754 and 740 residues, respectively. 
This explains why the mesophilic form has the larges radius of gyration, 
as shown in Fig.~\ref{fig:1}d. 

\begin{figure}[ht]
\begin{center}
\scalebox{0.68}{\includegraphics{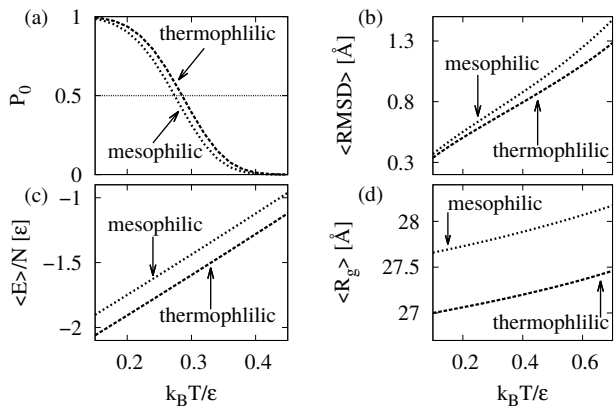}}
\caption{\label{fig:2} Similar to Fig.~\ref{fig:1} but for 
citrate synthase from thermophilic (2IBP) and mesophilic (1CTS) 
organisms in the open conformation. }
\end{center}
\end{figure}

We next compare the thermodynamic properties of CS dimers from 
thermophilic and mesophilic organisms in the open conformation 
(PDB codes 2IBP and 1CTS, respectively). 
As shown in Fig.~\ref{fig:2}a, the thermophilic CS 
exhibits larger $P_0 (T)$ and is thus
thermodynamically more stable than the mesophilic CS. 
The values of $T_f$  are
$0.285 \, \epsilon/k_B$ and $0.274 \, \epsilon/k_B$ 
for the thermophilic and mesophilic CS, respectively.
The difference is of order 10$^{\circ}$~C.
Fig.~\ref{fig:2}b shows that the mesophilic CS 
exhibits larger $\langle$RMSD$\rangle (T)$ 
and is thus more susceptible for thermal fluctuations 
than the thermophilic CS. 
A larger thermal stability of the thermophilic CS can be
also inferred from Fig.~\ref{fig:2}c which shows 
the internal energy per residue as a function of $T$.
We note also that the thermophilic CS is more compact than
than the mesophilic CS, see Fig.~\ref{fig:2}d. 
Overall, we observe that the qualitative differences between the 
proteins belonging to the organisms preferring high and low 
temperatures do not depend on whether the structures are
closed or open.

It is instructive to compare the open and closed CSs 
from the same organism (with PDB codes 1CTS and 2CTS, respectively). 
We note that the closed form is more compact than the open one. 
Its radius of gyration is smaller by 1.2~{\AA}.
The closed form exhibits a slower decrease of $P_0$ with $T$. 
As mentioned before, $T_f$ is $0.289 \; \epsilon /k_B$ for 
the closed form and $0.274 \; \epsilon /k_B$ for the open one. 
Since the two forms have identical sequences, the difference in $T_f$ 
should be related to the number of native contacts, which is 
2060 in the open conformation and 2138 in the closed conformation. 

The higher thermodynamic stability of the thermophilic proteins
is expected and required by the function. It is interesting to 
ask, however, what features of the native structure encode
this information. It has been shown recently, in a similar model,
that effective stiffness of virus capsids, as measured in
nanoindentation experiments, correlates with average coordination
number, $z$ \cite{virusplos}. Specifically, the corresponding
Young modulus is proportional to $(z \; - \; 6)^2$.
The average coordination number is defined as
\begin{equation}
z = 2 \frac{ \# {\rm native} \; {\rm contacts} \; + \; \# {\rm bonds} }{ \# {\rm residues} }
\end{equation}
and it describes how many neighbors, on an average, any residue has. 
It is determined in the native state  and the notion of neighborhood
relates to the dynamics of the system. Table~\ref{table:3} demonstrates that
the values of $T_f$ correlate with $z$ both in the open and closed conformation. 
The coordination number can be determined both for the dimers and 
individual monomers. The corresponding quantities $z_{\rm dimer}$ and $z_{\rm monomer}$
differ because there are contacts formed at the interface of the
two monomers. The number of the interfacial contacts in 1AJ8 is 289; 
in 2CTS it is 302, whereas in 1A59 it is only 266. 
For the closed forms the numbers are 322 for 2IBP and 252 for 1CTS.
Whichever quantity one takes, $z_{\rm dimer}$ or $z_{\rm monomer}$, 
the correlation with $T_f$ is evident, which indicates that 
both bulk and interfacial contacts contribute to the larger stability of 
the thermophilic forms. The larger value of $z$ is also consistent with the 
tighter packing of residues in thermophilic enzymes. 

\begin{table*}[ht]
\begin{center}
\begin{tabular}{|c|c|c|c|c|c|c|}
\hline
organism type & conformation & PDB code & $k_B T_f / \epsilon$ & $z_{\rm dimer}$ & $z_{\rm monomer}$ & $CO$ \\
\hline
\hline
thermophilic & closed & 1AJ8 & 0.303 & 7.31 & 6.54 & 0.0859 \\
\hline
mesophilic & closed & 2CTS &  0.289  & 7.01 & 6.33 & 0.0784 \\
\hline
cryophilic & closed & 1A59 & 0.287 & 6.99 & 6.29 & 0.0876 \\
\hline
\hline
thermophilic & open & 2IBP & 0.285 & 7.04 & 6.25 & 0.0846 \\
\hline
mesophilic & open & 1CTS & 0.274 & 6.71 & 6.14 & 0.0811 \\
\hline
\end{tabular}
\end{center}
\caption{\label{table:3} In both the open and closed conformation, 
$T_f$ correlates with the average coordination number 
both for CS dimers ($z_{\rm dimer}$) and monomers ($z_{\rm monomer}$). }
\end{table*}

It is interesting to ask whether the stability temperature 
is correlated with the average contact order
\begin{equation}
CO = \frac{1}{MN} \sum_{ij} \Delta_{ij}
\end{equation}
Here, $N$ is the number of residues in an amino acid chain, 
where $M$ is the total number of the native contacts, 
and $\Delta_{ij}=|i-j|$ is the sequence separation between 
residues $i$ and $j$ that form a native contact. 
Note that the contact order is uniquely defined only 
for single amino acid chains (monomers). 
The values of the average contact order as calculated from 
the contact maps of the five CS forms 
are given in the last column of Table~\ref{table:3}. 
Interestingly, we find no correlation between $CO$ and $T_f$. 
A similar conclusion has been reached in Ref.~\cite{Hoang_BJ_2003}.

\subsection{The temperature factors and RMSF}

\begin{figure}[ht]
\begin{center}
\scalebox{0.68}{\includegraphics{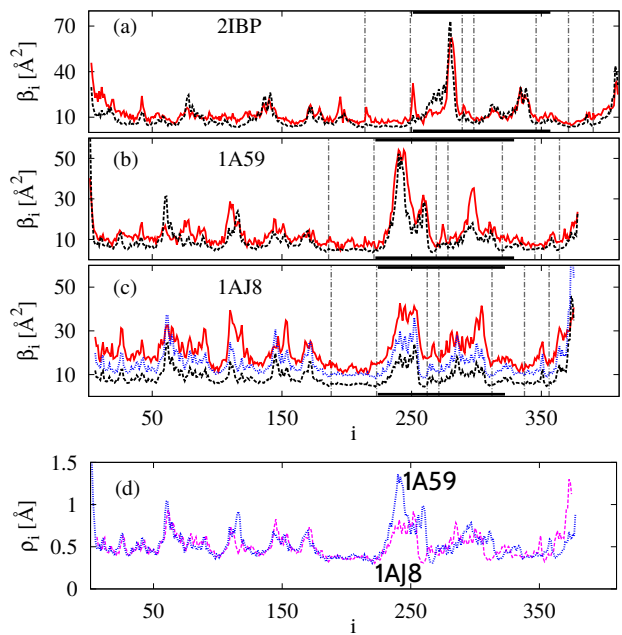}}
\caption{\label{fig:3} Temperature factors $\beta_i$ versus 
the residue number $i$ from X-ray crystallography experiments
(see the solid lines in red) and results of simulations
at $T = 0.3 \, \epsilon /k_B$ (see the dashed lines in black). 
The PDB codes of the different enzymes are given in figure panels. 
The vertical lines indicate the residues that form the active site 
as listed in Table~S1 in Supplementary Material \cite{SuppMat}.  
The thick horizontal lines show the location of the small domain. 
The blue dashed line in panel (c) shows the result of simulations 
at $T = 0.4 \, \epsilon /k_B$. 
Panel (d) shows RMSF along the sequence, $\delta_i$, computed at $T = T_f$ 
for 1AJ8 ($T_f = 0.303 \; \epsilon /k_B$; see the dashed line in magenta) and 
1A59 ($T_f = 0.287 \; \epsilon /k_B$; see the dotted line in blue). }
\end{center}
\end{figure}

To validate our simulations, we compare 
experimental and simulational temperature factors of 
the different CSs, see Fig.~\ref{fig:3}. The simulations have been
done at $T = 0.3 \, \epsilon/k_B$  which is close to $T_f$.
Although the temperature factors in a crystal may in general 
differ from those in a solution, 
this approach has been widely used to parametrize 
elastic network models \cite{Erman1997}. 

We obtain very good agreement for the 2IBP and 1A59 structures,
see panels (a) and (b) in Fig.~\ref{fig:3}. For 1AJ8, 
see Fig.~\ref{fig:3}(c), the simulation pattern closely resembles 
the distribution of the crystallographic $\beta$ factors 
but its magnitude is too small. However, simulations at 
$T = 0.4 \, \epsilon/k_B$ fit the experiment much better, 
see the blue line in Fig~\ref{fig:3}(c). 
We do not compare results for 1CTS because
the resolution of the mesophilic CS structure in 
the open conformation (1CTS) is only 2.7~{\AA}. 
The resolution for the structures of 2IBP, 1A59, and 1AJ8
is much better and is equal to
1.6~{\AA}, 2.1~{\AA}, and 1.9~{\AA} respectively.

We note that the average crystallographic $\beta$ factor,
$\bar{\beta} = \frac{1}{N} \sum_{i=1}^{N} \beta_i$, 
is $\bar{\beta}=21$~{\AA}$^2$ for the thermophilic CS in the closed conformation (1AJ8), 
and $\bar{\beta}=14$~{\AA}$^2$ for the cryophilic CS in the closed conformation (1A59). 
Importantly, in both cases the protein crystals have been reported to be obtained 
at the room temperature. 
The experimental data would thus imply that the thermophilic CS is, on average,  
more flexible than its cryophilic homologue at the same temperature. 
One would certainly expect the opposite relation. 
This inconsistency might be a reason for the quantitative disagreement 
between simulations and experiment in the case of 1AJ8, 
see Fig.~\ref{fig:3}(c).

In all of the analyzed structures, the residues that have 
the largest $\beta$ factors are localized in the small domain 
(the region highlighted in Fig.~\ref{fig:3}), primarily
in the segments that bind the CoA molecule. 
In contrast, the residues that bind citrate 
(indicated by the vertical lines in Fig.~\ref{fig:3}) 
have relatively small temperature factors. 

When one compares the $\beta$ factors of the thermophilic
and cryophilic proteins in the closed conformation, 
see panels (b) and (c) of Fig.~\ref{fig:3}, 
then they are seen to map out very similar looking patterns
no matter whether one uses the experimental results at the room temperature 
(red solid lines in Fig.~\ref{fig:3}) or the simulational results 
at $T=0.3 \; \epsilon /k_B$ (black dashed lines in Fig.~\ref{fig:3}). 
When $\delta_i$ or $\beta_i$ are calculated at a higher
$T$ for the thermophilic form and at a lower $T$ for the 
cryophilic form then the match in the patterns is better quantitatively. 
Fig.~\ref{fig:3}(d) compares $\delta_i$ for 1AJ8 and 1A59 at their respective 
stability temperatures. The amplitudes of the fluctuations are seen to be comparable.

\subsection{Fluctuations of the active site}

The citrate binding site comprises seven evolutionarily conserved amino acids: 
three histidine residues, three arginine residues and one aspartic acid 
\cite{Karpusas1990,Bell2002,Bjelic2008}. 
They are depicted in Fig.~\ref{fig:totable2} and summarized in 
Table~S1 which is provided in Supplementary Material \cite{SuppMat}.   
Three of them, which we denote here as His$_2$, His$_3$ and Asp$_1$, 
are directly involved in the chemical reaction 
that results in citrate formation \cite{Karpusas1990}. 
They are highlighted in Table~S1 \cite{SuppMat} in bold. 
His$_{3}$, Arg$_{1}$ and Aps$_{1}$ reside in the small domain. 
His$_{1}$, His$_{2}$, Arg$_{2}$ and Arg$_{3}$ are located in the large domain 
(His$_{2}$ is at the N-terminal end of the large domain). 
Arg$_{3}$ is the only citrate-binding residue that resides on the other monomer. 

\begin{figure}[ht]
\begin{center}
\scalebox{0.22}{\includegraphics{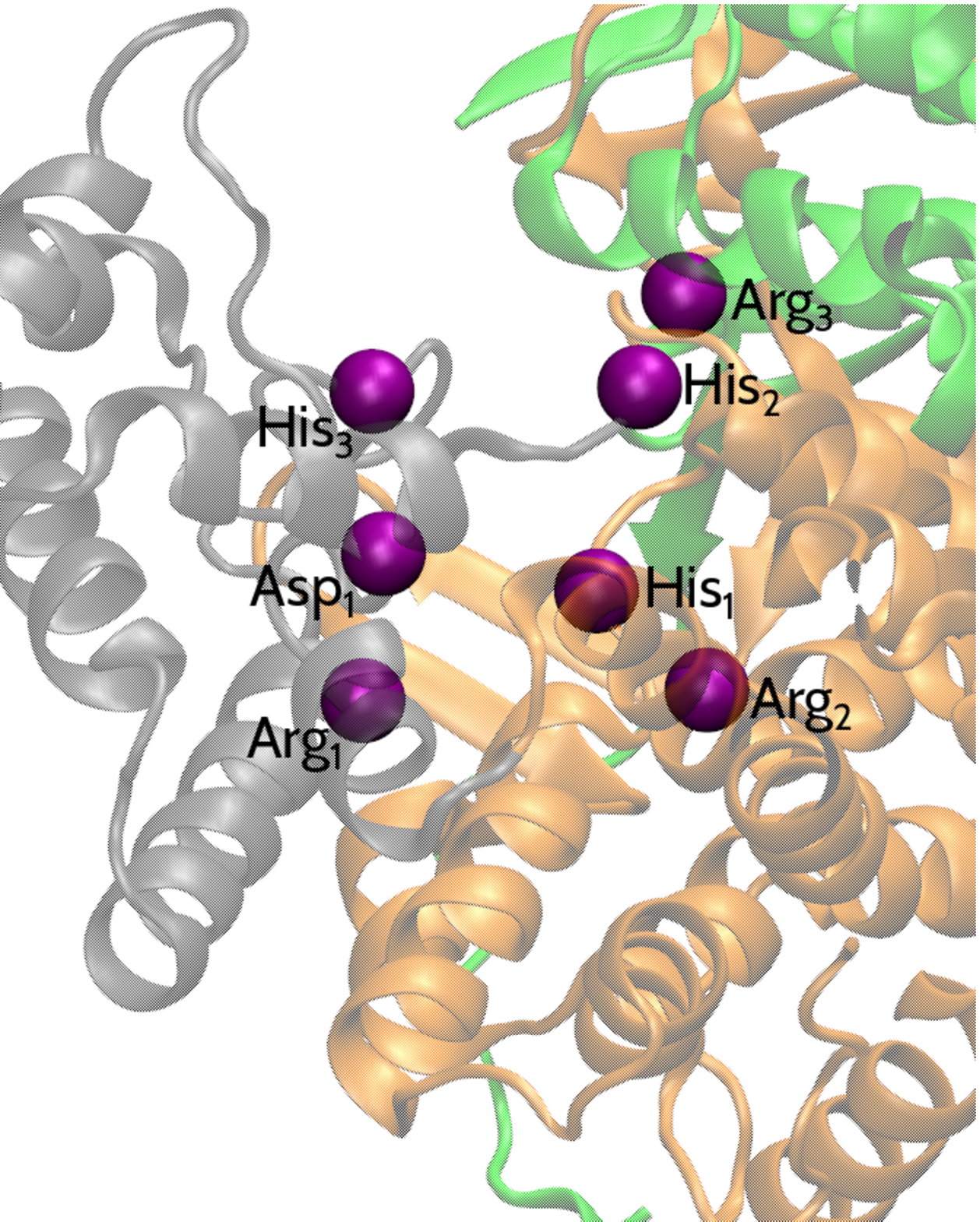}}
\scalebox{0.22}{\includegraphics{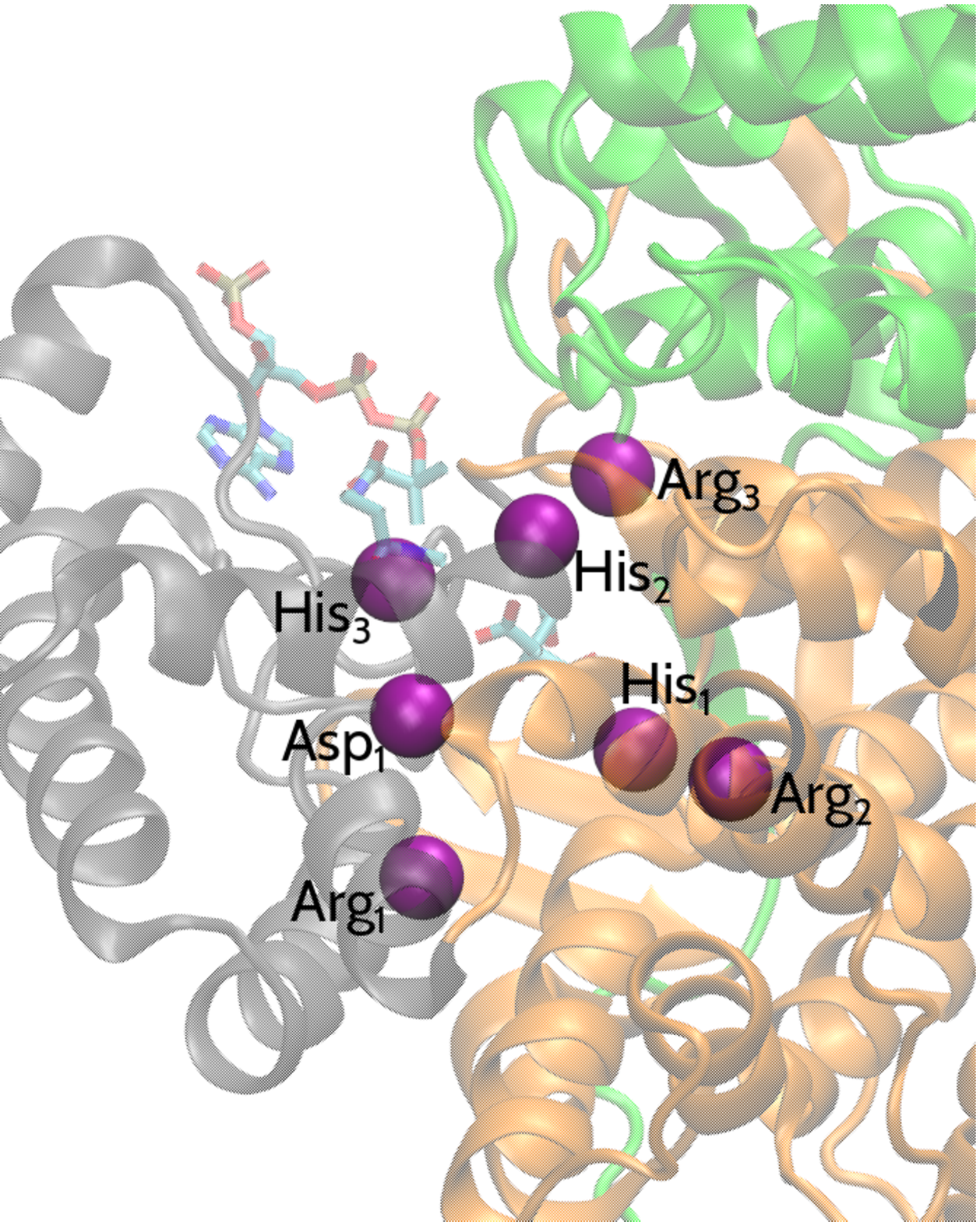}}
\caption{\label{fig:totable2} Structures of the citrate binding site 
in the open (2IBP, left) and closed (1AJ8, right) conformations. 
The seven evolutionarily conserved residues that are involved in binding 
the citrate molecule are shown as purple spheres. 
The citrate and CoA molecules are shown in the stick representation. 
The color code in the same as in Fig.~\ref{fig:0}. }
\end{center}
\end{figure}

\begin{figure}[ht]
\begin{center}
\scalebox{0.42}{\includegraphics{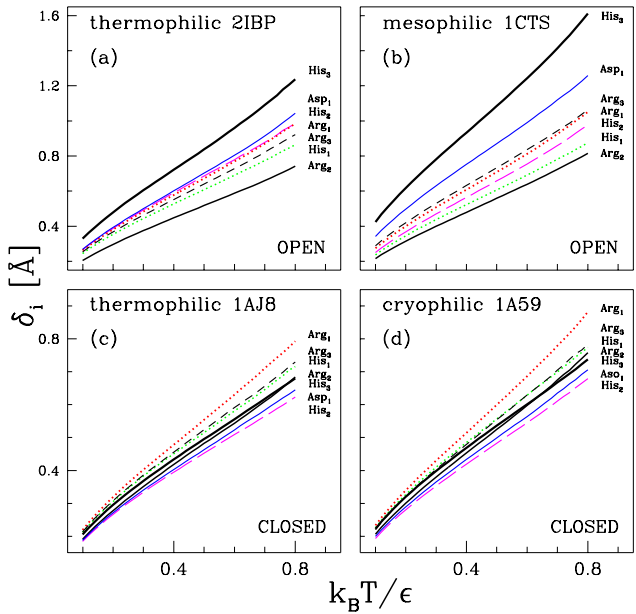}}
\caption{\label{fig:4} RMSF $\delta_i$ of the seven 
residues forming the active site as a function of $T$. }
\end{center}
\end{figure}

In Fig.~\ref{fig:3} the citrate binding residues are labeled 
with vertical lines. All of them exhibit relatively small 
spatial fluctuations, as indicated by low $\beta$ factors. 
Fig.~\ref{fig:4} shows the RMSFs of 
these seven amino acid residues as a function of $T$. 
In the closed conformation, the active site in the thermophilic and 
cryophilic enzymes fluctuate in the same manner 
although, for any fixed $T$, the magnitude of fluctuations is somewhat larger  
in the cryophilic CS, see panels c and d in Fig.~\ref{fig:4}. 
The residues that exhibit largest RMSFs are Arg$_1$, Arg$_3$ and His$_1$. 
The residues that fluctuate most weakly are His$_2$, Asp$_1$ and His$_3$, 
which happen to be the three amino acids that are directly 
involved in the chemical reaction of citrate formation. 

Our results at temperatures between 0.35 and 0.45~$\epsilon/k_B$ 
agree qualitatively with the results of 
extensive all-atom molecular dynamics simulations \cite{Bjelic2008}, 
which show that at $T=300$~K the RMSFs of the active site residues 
in contact with the citrate molecule 
are in the range between 0.4 and 0.6~{\AA}, 
depending on the organism (thermophilic, mesophilic or cryophilic) 
and boundary conditions (periodic or spherical). 

In the open conformation, the active site in the thermophilic and 
mesophilic enzymes fluctuate in a similar manner but quite differently 
than in the closed conformation, see panels a and b in Fig~\ref{fig:4}. 
Here, the residues that fluctuate the most are His$_3$ and Asp$_1$, 
which are involved in the reaction of citrate formation.
The residues that exhibit smallest fluctuations are Arg$_2$ and His$_1$. 
Also the magnitude of RMSFs is larger than in the closed conformation. 

Interestingly, the three residues involved directly in the catalytic activity, 
His$_3$, His$_2$ and Asp$_1$, exhibit rather small thermal fluctuations 
in the closed conformation, whereas in the open conformation the thermal 
fluctuations of these three residues are significantly enhanced. 
This result seems consistent with the division of functions 
between residues in the active site: 
some amino acids are tailored to binding the substrate, 
whereas others participate in the catalytic process. 

\subsection{Collective motions of the active site}

\begin{figure}[ht]
\begin{center}
\scalebox{0.65}{\includegraphics{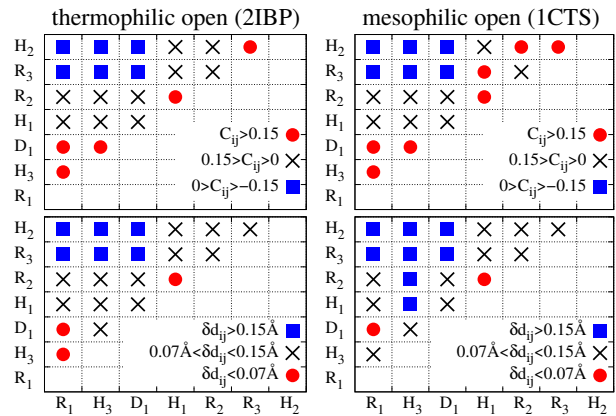}}
\caption{Equal-time correlation coefficients (upper panels) 
and variation of distances (lower panels) 
between the seven residues forming the active site in the open conformation. 
The results were obtained in simulations of the 
thermophilic (left panels) and mesophilic (right panels) CS dimers
at $T = 0.3 \, \epsilon/k_B$. 
The largest positive correlations and the smallest distance variations are shown as red circles. 
The negative correlations and the largest distance variations are shown as blue squares. 
The intermediate cases are indicated by black crosses. }
\label{fig:5}
\end{center}
\end{figure}

In order to identify collective motions of the active site residues, 
we compute the equal-time correlation coefficients $C_{ij}$ 
as defined in Eq.~(\ref{eq:Cij}) for all pairs $(i,j)$ 
of the residues forming the active site. 
The upper panels of Fig.~\ref{fig:5} show correlation levels 
of these pairs in the thermophilic (left panel) and 
mesophilic (right panel) enzymes in the open conformation. 
The positive correlations with $C_{ij}>0.15$ are labeled 
with red circles and indicate that the pairs of residues 
often move together in the same direction. 
The negative correlations are depicted as blue squares and indicate that 
the pairs have a tendency to move simultaneously in opposite directions. 
The results shown in Fig.~\ref{fig:5} were obtained in simulations at 
temperature $T = 0.3 \, \epsilon/k_B$ but we checked that the same 
correlation pattern persists at temperatures 
between $0.2 \; \epsilon/k_B$ and $0.5 \; \epsilon/k_B$. 

Interestingly, all the active site residues that are located in the small domain, 
His$_{3}$, Arg$_1$ and Asp$_{1}$, appear to move together as a unit, 
which is indicated by significant positive correlations for all the three residue pairs. 
The three residues in the small domain exhibit motions that are anti-correlated with 
the motions of His$_{2}$ and Arg$_{3}$, which are placed in the large domain. 
Therefore, the collective motions of the active site in the open conformation 
seem to be due to the displacements of the small domain relative to the large domain. 

To further quantify this observation we compute the variations of 
the inter-residue distances, 
$\delta d_{ij} =\langle (d_{ij} - \langle d_{ij} \rangle)^2 \rangle^{1/2}$, 
where $\langle d_{ij} \rangle$ is the average distance between residues 
$i$ and $j$ forming the active site. 
The lower panels of Fig.~\ref{fig:5} illustrate the distance variations 
in the open conformation of the thermophilic (left panel) and 
mesophilic (right panel) enzymes. 
The smallest distance variations (red circles) are observed for 
the residue pairs that exhibit correlated motions with $C_{ij}>0.15$. 
The largest distance variations (blue squares) are observed almost 
exclusively for the residue pairs that exhibit negative spatial correlations 
and, thus, have the propensity to move in opposite directions. 
Importantly, these results show that the motions of largest amplitudes 
occur primarily between the residues located in different domains, 
and the small-amplitude motions are mainly between the residues residing 
in the same domains. 

In the open conformation of the thermophilic CS, we can distinguish 
three groups of spatially-correlated residues: 
the first group comprises the three residues in the small domain, 
Arg$_{1}$, His$_{3}$ and Asp$_{1}$;
the second group consists of His$_{1}$ and Arg$_{2}$;
and the third group is His$_{2}$ and Arg$_{3}$. 
Within each of the groups, the individual residues show a propensity 
to move jointly with other group members. 
For example, His$_{1}$ and Arg$_{2}$ often move together 
in the same direction ($C_{ij}=0.26$) but almost independently 
from other residues of the active site. 
Interestingly, all the residues in the first group are spatially 
anti-correlated with the residues in the third group. 
These two groups can be thus visualized as the lower and upper 
'jaws' moving in an up-and-down manner. 

In the open conformation of the mesophilic CS, we observe more correlations 
between the active site residues present in the large domain. 
Here, the 'upper jaw' (His$_{2}$ and Arg$_{3}$) appears to be connected 
more stiffly to the 'hinge' (His$_{1}$ and Arg$_{2}$) but performs even 
larger motions ($\delta d_{ij} > 0.2$~{\AA}) 
against the 'lower jaw' (Arg$_{1}$, His$_{3}$ and Asp$_{1}$).

In the closed conformation, both in the thermophilic and cryophilic enzymes, 
we observe only positive correlations, $C_{ij}>0$, and rather small 
inter-residue distance variations, $\delta d_{ij} < 0.07$~{\AA}, 
see Fig.~S1 in Supplementary Material \cite{SuppMat}.  
Here, due to the presence of the citrate and CoA molecules, 
the small domain is tightly bound to the large domain, and 
the active site residues have no freedom to perform any large-amplitude motions. 
Therefore, the residues in the small domain are positively correlated with 
those in the large domain. Interestingly, the active sites in 
the thermophilic and cryophilic CS exhibit the same pattern of correlations, 
see the top panels in Fig.~S1.

\subsection{Kinetics of thermal unfolding and dissociation \label{sec:3.5}}

Another way to assess thermal stability of proteins is to simulate their 
unfolding at elevated temperatures as analyzed theoretically 
in Ref.~\cite{Sulkowska_JCP_2005}.    
The unfolding simulations start at the native state and finish when 
all nonlocal contacts get broken, which defines the unfolding time $t_{\rm unf}$. 
Specifically, the nonlocality refers to the sequential distance 
$|i-j|>l$. We take $l=10$ in this study. 
In Ref.~\cite{Sulkowska_JCP_2005}, $l$ has been taken to be 4 to eliminate 
the local contacts in $\alpha$-helices, but this choice is too demanding 
computationally in the current context. 

We performed unfolding simulations at high temperatures only for 
the CS dimers in the open conformation because dissociation of 
the citrate and CoA molecules from CS occurs much faster than protein unfolding or 
dimer dissociation (kinetic constants for the successive steps of the CS reaction 
at physiological conditions are detailed, for example, in Ref.~\cite{plos_one_2008}). 
At any given $T$ between $1.1 \; \epsilon / k_B$ and $1.7 \; \epsilon / k_B$, 
we run $200$ trajectories. 
We checked that $t_{\rm unf} < 25 000 \; \tau$ in 
all the simulation trajectories. 

\begin{figure}[ht]
\begin{center}
\scalebox{0.65}{\includegraphics{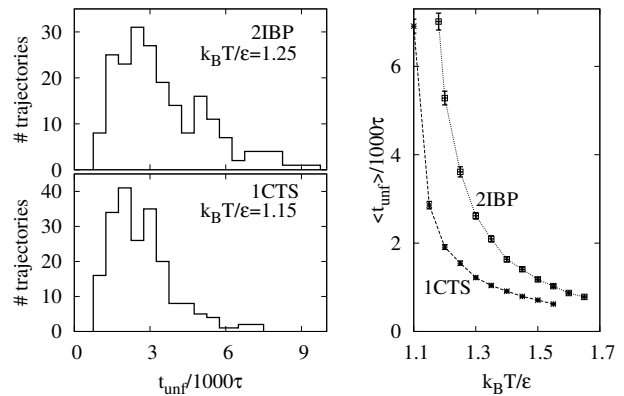}}
\caption{\label{fig:7} 
Thermal unfolding of the thermophilic (2IBP) and mesophilic (1CTS) CS 
in the open conformation. The panels on the left hand side show histograms of 
the unfolding time, $t_{\rm unf}$, at temperatures $T=1.25 \; \epsilon /k_B$ 
(2IBP, upper-left panel) and $T=1.15 \; \epsilon /k_B$ (1CTS, lower-left panel). 
The panel on the right hand side shows the average unfolding time 
as a function of temperature. 
} 
\end{center}
\end{figure}

The histograms of $t_{\rm unf}$ for the thermophilic (2IBP) and 
mesophilic (1CTS) CS in the open conformation are shown in the left-hand-side 
panels of Fig.~\ref{fig:7}.  These histograms demonstrate that 
the thermophilic CS unfolds slower at $T=1.25 \; \epsilon /k_B$ than 
the mesophilic CS at $T=1.15 \; \epsilon /k_B$, which means that 
the thermophilic enzyme is more resistant to thermal denaturation. 
Also, the histograms seem to display multi-peak profiles, which suggests 
that there might be more than one characteristic time scale governing 
the thermal unfolding of the CS dimers. 
To further characterize the unfolding kinetics, it is instructive to calculate 
the fraction of the number of trajectories, $\phi_f (t)$, in which the unfolding event 
has not occurred within time $t$ from the beginning of the simulation. 
The fraction $\phi_f (t)$ provides an estimate for the probability of 
not unfolding CS within time $t$. As can be seen in Fig.~S2 \cite{SuppMat}, 
there is an initial lag phase in which $\phi_f (t) = 1$. 
The lag phase is followed by a fast decrease of $\phi_f (t)$. 
Interestinlgy, we find that the decay of $\phi_f$ with $t$ is not exponential. 


The right-hand-side panel of Fig.~\ref{fig:7} shows the average unfolding time, 
$\langle t_{\rm unf} \rangle$, as a function of $T$. 
The average is taken over 200 trajectories. 
The error bars correspond to the standard error of the mean. 
At any specified temperature, the thermophilic CS (2IBP) exhibits 
significantly larger unfolding time than the mesophilic CS (1CTS). 
This result shows that the thermophilic enzyme is thermally more stable 
than its mesophilic homologue. Based on the dependence of $\langle t_{\rm unf} \rangle$ 
on $T$ as shown in Fig.~\ref{fig:7}, the difference in stability temperatures 
is of the order $0.1 \; \epsilon /k_B$, which corresponds to about 80$^{\circ}$~C. 

We use an analogous method to quantify thermal dissociation of CS dimers. 
The dynamics of the system is exactly the same as in the unfolding 
simulations but only the contacts between the monomers are monitored. 
The simulations start at the native state and finish when all contacts 
between the monomers get broken, which defines the unbinding time $t_{\rm unb}$. 
In the range of temperatures studied, 
$t_{\rm unb} < 25 000 \; \tau$ in all the trajectories. 

\begin{figure}[ht]
\begin{center}
\scalebox{0.65}{\includegraphics{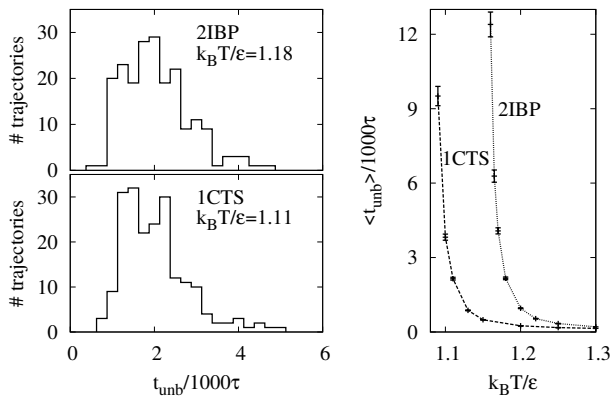}}
\caption{\label{fig:8} 
Thermal dissociation of the thermophilic (2IBP) and mesophilic (1CTS) CS 
dimers in the open conformation. The left-hand-side panels show histograms of 
the unbinding times at $T=1.11 \; \epsilon /k_B$ (2IBP) and 
$T=1.18 \; \epsilon /k_B$ (1CTS). The right-hand-side panel shows 
$\langle t_{\rm unf} \rangle$ as a function of $T$. 
} 
\end{center}
\end{figure}

The left panels of Fig.~\ref{fig:8} show histograms of the unbinding time 
for the thermophilic (2IBP) and mesophilic (1CTS) CS in the open conformation 
at $T=1.11 \; \epsilon /k_B$ and $T=1.18 \; \epsilon /k_B$, respectively. 
These data show that the thermophilic dimer is more resistant to 
thermal dissociation. To characterize the dissociation kinetics, 
we compute the fraction of the number of trajectories, $\phi_b (t)$, in which 
the unbinding event has not occurred within time $t$. 
Fig.~S2 \cite{SuppMat} shows $\phi_b (t)$ for 2IBP and 1CTS at different temperatures. 
We note that the decay of $\phi_b$ with $t$ is not exponential. 

The right panel of Fig.~\ref{fig:8} shows the average unbinding time, 
$\langle t_{\rm unb} \rangle$, as a function of $T$. 
The average is taken over 200 trajectories, and the error bars correspond to 
the standard error of the mean. At any given temperature, $\langle t_{\rm unb} \rangle$ 
of the thermophilic CS dimer (2IBP) is significantly larger than that of 
the mesophilic enzyme (1CTS). Interestingly, the two plots 
of $\langle t_{\rm unb} \rangle$ versus $T$ overlap if the temperature is shifted by 
about $0.07 \; \epsilon /k_B$. This temperature shift corresponds roughly 
to 60$^{\circ}$~C, which is consistent with the difference in 
the optimal growth temperatures, see Table~\ref{table:1}.

\section{Summary}

We performed Langevin dynamics simulations of CS dimers from thermophilic, 
mesophilic and cryophilic organisms using a coarse-grained structure-based model 
that does not differentiate between the ionic, hydrophobic or van der Waals 
interactions. In fact, the interactions between different amino acid residues 
are described in this model by the same potential energy function. Although 
this model might seem oversimplified, it correctly predicts that 
the thermophilic CS is thermally more stable than the mesophilic and cryophilic ones. 
It also yields root-mean-square fluctuates of single amino acid residues 
that are fully consistent with crystallographic temperature factors. 
Therefore, the precise location of the amino acid contacts appears 
to be a key element in explaining thermodynamic properties and 
local flexibility of enzymes.

The difference of the thermodynamical stability temperatures, $\Delta T_f$, 
can be obtained from the native state probabilities (see section \ref{sec:3.1}). 
For the CS dimers from thermophilic and cryophilic organisms 
in the closed conformation we obtain $\Delta T_f \approx 14^{\circ}$~C; 
for the thermophilic and mesophilic CS dimers
in the open conformation we get $\Delta T_f \approx 10^{\circ}$~C. 
These values are smaller than expected. 
The expected difference in melting temperatures between 
the mesophilic and thermophilic CS dimers is about $20^{\circ}$~C 
as reported in a recent study~\cite{Pfleger_2013}. 
The likely reason for the fact that our model underestimates the transition 
temperature differences is that it lacks some relevant sequence effects. 
As discussed in section~\ref{section:Methods}, the atomic structures of the CS proteins 
are used in the Go-type model only to construct the contact maps whereas 
the chemical composition of the proteins is not directly included 
in the energy function of the model as given by Eq.~(\ref{eq:Hamiltonian}). 

The temperature of optimal folding, $T_f$, is one measure of 
the thermodynamic stability of proteins.
Another measure is provided by the kinetics of thermal unfolding 
(see section \ref{sec:3.5}). 
These two measures are quite different as they refer to equilibrium properties 
near the native state and dynamic properties away from the native state, respectively.
One measures the frequency of the situations in which all contacts are present 
simultaneously, and the other focuses on rupture events up to a given sequential length. 
The stability temperature difference 
obtained from the thermal unfolding of CS dimers from thermophilic and 
mesophilic organisms (see Fig.~\ref{fig:7}) is of the order of 80$^{\circ}$~C. 
We note that this value can be affected by the partition of the native 
contacts into local and nonlocal (we used the sequential 
distance cut-off $l=10$ in our analysis). On the other hand, 
the stability temperature difference obtained from the thermal 
dissociation of CS dimers from thermophilic and mesophilic organisms 
is about 60$^{\circ}$~C (see Fig.~\ref{fig:8}). 
We note that this value is unaffected by the choice of the sequential 
distance cut-off, $l$, as we take into account all inter-monomer 
contacts in the analysis of the simulation data. 
Interestingly, this temperature difference, 60$^{\circ}$~C, 
is consistent with the difference in the optimal growth temperatures 
as given in Table~\ref{table:1}. 

We analyzed thermal fluctuates and collective motions of the active site 
in CS enzymes from different organisms, both in the open and closed conformations. 
We find that the three residues that are directly involved in the chemical 
reaction of citrate formation exhibit rather small fluctuations in 
the closed conformation. In the open conformation, however, the thermal 
fluctuations of these three residues are significantly enhanced. 
We also find that the collective motions of the active site in the open conformation 
are mainly due to the movements of the small domain relative to the large domain. 
In the open conformation, due to the presence of the citrate and CoA molecules, 
the small domain is tightly bound to the large domain, and the active site 
residues have no freedom to perform any large-amplitude collective motions. 
Taken together, our results show that the relatively simple structure-based model
correctly captures the similarities and differences in thermodynamic
and kinetic properties of the different forms of the CS enzyme.

\vspace{0.5cm}

\noindent {\bf Acknowledgments}

\noindent This work has been supported by the Polish National Science Center grant 
No. 2012/05/B/NZ1/00631 (BR) and 
by the ERA-NET grant FiberFuel (MC).

\end{document}